\documentclass[10pt, conference, compsocconf]{IEEEtran} 

\usepackage{amsmath}
\usepackage{amssymb}
\usepackage{graphicx}
\usepackage{cite}
\usepackage{subfigure}
\usepackage{float}
\usepackage{url}       
\usepackage{supertabular}
\usepackage{exscale}
\usepackage{array}
\usepackage{underscore}
\usepackage{verbatim}
\usepackage{multirow}
\usepackage{rotating}
\usepackage{boxedminipage}
\usepackage{colortbl}
\usepackage{color}
\usepackage{listings}

\def\ten#1{\texttt{#1}}
\def\ien#1{\textit{#1}}
\newcommand{\BibTeX}{{\rm B\kern-.05em{\sc i\kern-.025em b}\kern-.08em
    T\kern-.1667em\lower.7ex\hbox{E}\kern-.125emX}}

\begin{document}

\lstset{language=VHDL,
  basicstyle=\scriptsize\ttfamily,
  keywordstyle=\color{blue}\bfseries,
  identifierstyle=\bfseries,
  commentstyle=\color{green}\itshape,
  stringstyle=\bfseries\ttfamily,
  showstringspaces=false,
  numbers=none,
  frame=single,
  frameround=tttt
} 

\title{Design space exploration tools for the {ByoRISC} configurable processor family}

\author{\authorblockN{Nikolaos Kavvadias}
\authorblockA{Independent Research Scientist\\
Kornarou 12, 35100 Lamia, Greece\\
Email: nikos@nkavvadias.com}
\and
\authorblockN{Spiridon Nikolaidis}
\authorblockA{Department of Physics\\
Aristotle University of Thessaloniki\\
54124 Thessaloniki, Greece\\
Email: snikolaid@physics.auth.gr}}

\maketitle

\begin{abstract}
In this paper, the {ByoRISC} (``Build your own {RISC}'') configurable application-specific instruction-set processor (ASIP) family is presented. {ByoRISCs}, as vendor-independent cores, provide extensive architectural parameters over a baseline processor, which can be customized by application-specific hardware extensions (ASHEs). Such extensions realize multi-input multi-output (MIMO) custom instructions with local state and load/store accesses to the data memory. ByoRISCs incorporate a true multi-port register file, zero-overhead custom instruction decoding, and scalable data forwarding mechanisms. Given these design decisions, {ByoRISCs} provide a unique combination of features that allow their use as architectural testbeds and the seamless and rapid development of new high-performance ASIPs.   

The performance characteristics of ByoRISCs, implemented as vendor-independent cores, have been evaluated for both ASIC and FPGA implementations, and it is proved that they provide a viable solution in FPGA-based system-on-a-chip design. A case study of an image processing pipeline is also presented to highlight the process of utilizing a {ByoRISC} custom processor. A peak performance speedup of up to 8.5$\times$ can be observed, whereas an average performance speedup of 4.4$\times$ on Xilinx Virtex-4 targets is achieved. In addition, {ByoRISC} outperforms an experimental VLIW architecture named {VEX} even in its 16-wide configuration for a number of data-intensive application kernels.
\end{abstract}


\section{Introduction and related work}
\label{sec:intro}
Contemporary embedded system design involves the use of configurable and extensible processor cores \cite{Gonzalez00}
such as Xilinx MicroBlaze\footnote{\url{http://www.xilinx.com}}, 
and Altera Nios-II\footnote{\url{http://www.altera.com/products/ip/processors/nios2/}}, offering architecture customization 
possibilities. Configurability lies in tuning architectural parameters, while extensibility usually refers 
either to tightly-coupled modifications obtained by adding single-, multi-cycle or pipelined versions of 
custom functional units or by loosely-coupled accelerators not directly integrated within the processor pipeline. 
Recent work by \cite{Sirowy07} advocates that both the custom instruction (CI) and coprocessor 
approaches should be considered simultaneously by formalizing the problem as a 
form of two-level partitioning. 

A disciplined approach to CI generation for extensible processors is found in \cite{Goodwin03} 
where the Xtensa processor\footnote{\url{http://www.tensilica.com}} is augmented with CIs 
that may combine VLIW, SIMD or fused (chained) operations. Although the Xtensa framework is 
highly automated and feature-rich, simultaneous generation of disjoint optimal MIMO CIs is 
not considered; instead the CI generation process is divided in distinct stages with different 
objectives. CI generation is also used for designing custom coprocessors 
(ARM OptimoDE \cite{ClarkN05}) or two-input/one-output functional units (MIPS CorExtend \cite{Halfill03}) with internal 
register storage \cite{Leupers06}.

MOLEN \cite{Vassiliadis04} is a relevant approach that extends a basic architecture (PowerPC) with new 
instructions to interface and configure a number of loosely-coupled 
custom computation units. While MOLEN permits the simultaneous operation of the processor core and 
these units, it is not usable for optimizing fine-grain program regions. For the specific 
coprocessor paradigm used, the control/data communication overhead often prohibits the implementation of 
useful extensions for irregular code. 

Similarly to MOLEN, the Xilinx MicroBlaze vendor-specific core follows the coprocessor 
paradigm by using a communication interface named FSL (Fast Simplex Link) \cite{Rosinger04}.
Microblaze uses special ``put'' and ``get'' instructions to exchange control and data over 
a FIFO interface among the processor core and the extension units. Again, for establishing the 
concurrent operation of both the core and these units, extensive design considerations are 
required by the designer. This approach is also only suitable for accelerating coarse-grain program 
regions. 

Nios-II is a soft processor that stands closer to the ByoRISC approach. Nios-II 
provides a well-defined tightly-coupled interface to CI units, embedded within 
the processor pipeline. A specific opcode is reserved in the Nios-II instruction-set 
architecture for enabling these operations. However, this approach is only an 
incremental enhancement to the base Nios-II architecture and suffers from 
problems that arise when significant performance acceleration has to be achieved: 
a) legacy instruction encodings pose significant limitations, 
and b) CIs are limited to two read and one write programmer-visible operands.

Recent work in \cite{Chen06} allows the extension of a processor by a unit 
that can execute custom functionalities with up to six inputs and three 
outputs. This approach steps forward from the Nios-II limitation, however it 
is still affected by the limited macroinstruction encoding space; ByoRISC overcomes this 
problem by the usage of an intrinsic decoding phase. 

The ByoRISC architecture overcomes many of the aforementioned 
problems \cite{Kavvadias08b}. {ByoRISCs} enable the use of large MIMO operation clusters (typically with up to 
8 input and 8 output operands) by utilizing a configurable multi-port register file 
and without negatively affecting instruction encodings and the runtime behavior of the 
instruction decoder. An additional pipeline stage is required for the predecoding of 
register operands used by CIs. A scalable data forwarding architecture eliminates all hardware 
interlocks, and allows the close scheduling and issue of successive CIs. Also, the 
pipeline can be extended via multiple execution stages, at design configuration 
time, if demanded by the mapping of MIMO CIs to pipelined ASHEs. Different sets of 
CIs can be configured at different times.
These features combined make the service of arbitrary zero-overhead CIs within the 
processor a unique research testbed for the development of ASIPs that can be 
mapped to both ASIC and FPGA platforms. 

Overall, this work establishes the important contribution of a novel processor family named {ByoRISC}\footnote{A binary release of the software development (without the \ien{gcc} port) and simulation toolkit for ByoRISC is available: \url{http://www.nkavvadias.com/misc/byorisc-demo-0.0.1.zip}} that aims to serve as a vendor-independent infrastructure for the development of ASIPs. ByoRISC provides a clean, 
orthogonal architecture for architectural experimentation on future ASIPs for data-intensive processing tasks 
with the help of an assisting design space exloration tool, named YARDstick \cite{Kavvadias07}. 

The rest of this paper is organized as follows. The {ByoRISC} architecture is presented in detail in Section 2. Section 3 discusses custom instruction generation using the YARDstick DSE (Design Space Exploration) tool. Section 4 discusses area and timing characterization of ASIC and FPGA implementations of {ByoRISC} processors. In Section 5, a {ByoRISC}-based system is used to accelerate an image processing application set and is compared to a parameterized VLIW architecture. Finally, Section 6 summarizes the paper.

\section{The {ByoRISC} architecture}
\label{sec:byorisc}

A key issue for the success of a SoC design involving ASIPs is the ease of application development for the corresponding platform. For fully supporting high-level compiled languages, the ASIP has to provide a self-contained set of primitive operators. For example, the instruction set of the SABRE RISC processor includes 28 integer instructions to fully support the ANSI C integer subset \cite{SABRE}; the same concept applies to industrial architectures such as MicroBlaze, Nios-II and MIPS32 modern embedded soft processors. The need for a fundamental RISC instruction set implies the development of an underlying architecture that ought to be common across processor variations in order to sustain code reuse, minimal application compatibility requirements and tool stability. 

A proper instruction set partitioning for a customizable processor family would define base, coprocessor and custom subsets. The base instruction set is comprised of primitive instructions that ought to be supported across all processor variants as well as derived instructions that can be directly implemented in hardware, otherwise they should be emulated by embedded software.

\subsection{Overview of the {ByoRISC} application-customizable processors}
\label{subsec:byorisc:overview}

The {ByoRISC} architecture encompasses the following characteristics:

\begin{itemize} 
\item{32-bit instruction and data word length; cacheless Harvard memory architecture.}
\item{Base instruction set comprising of 22 primitive and 22 derived instructions.}
\item{64--256 distinct primary opcodes, up to 192 available to CI extensions.}
\item{Configurable number of execution pipeline stages. The total number of pipeline stages 
of ByoRISCs is 5 (minimum), 6 (supporting CIs) or 5+ (multiple execution stages with CIs).}
\item{Optional support for the ZOLC (Zero-Overhead Loop Controller) architecture \cite{Kavvadias08} for the elimination of looping overheads within nested loop structures of arbitrary complexity. \footnote{ZOLC is supported in the ByoRISC ArchC simulator and the XiRisc VHDL model \cite{Campi01, Kavvadias08}; it has not been integrated in the ByoRISC VHDL model.}}
\item{Register file size can be configured from a minimum of 16 to a maximum of 256 entries.}
\item{Configurable number of read (2--8) and write (1--8) register file ports.}
\item{Interface specifications for incorporating tightly-coupled and local coprocessor application-specific hardware extensions.
\footnote{Coprocessor interfacing uses the SimpCon specification (\url{http://opencores.org/project,simpcon}); 
it is under development.}}
\item{Designed to be used with synchronous read RAM storage for instructions, data and CI 
predecoding information.}
\end{itemize}

\begin{figure}[htb]
  \centering
  \includegraphics[width=0.4375\textwidth]{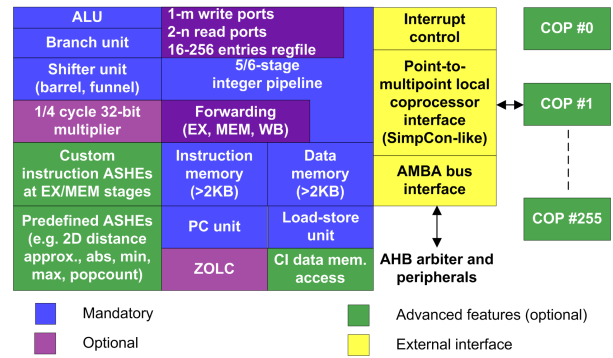}
  \caption{Conceptual diagram of the {ByoRISC} architecture.}
  \label{Fig:byorisc:cdiag}
\end{figure}

A conceptual diagram of the {ByoRISC} architecture highlighting its constituent components is shown in Fig.~\ref{Fig:byorisc:cdiag}.

\subsection{Instruction formats}
\label{subsec:byorisc:if}

\begin{figure}[htb]
  \centering
  \includegraphics[width=0.4375\textwidth]{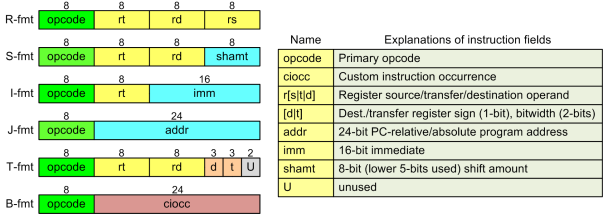}
  \caption{{ByoRISC} instruction formats.}
  \label{Fig:byorisc:if}
\end{figure}

The {ByoRISC} instruction formats (Fig.~\ref{Fig:byorisc:if}) have been designed for maximum orthogonality in order to simplify the instruction decoding hardware. For this reason, the instruction fields for all formats start at an 8-bit (byte) boundary, with having only the type conversion instruction (cvt) subdividing its secondary opcode to subfields for specifying sign and bitwidth of source/destination operands.

There are five distinct formats in the base instruction repertoire: \ten{R-fmt} for instructions with two source and one destination register operands, \ten{S-fmt} for shifts by an immediate constant, \ten{I-fmt} for accessing 16-bit immediates, \ten{J-fmt} for jump instructions and \ten{T-fmt} for type conversion operations. Coprocessor instructions derived from the MIPS-I/32 specification follow the \ten{S-fmt} while CIs are encoded in the \ten{B-fmt}. The {\it ciocc} field denotes a specific occurrence of a CI usage in ASIP-targeted applications. 

\subsection{The {ByoRISC} instruction set}
\label{subsec:byorisc:isa}

The {ByoRISC} instruction set shares characteristics to typical load-store machines such as DLX and MIPS-I/32. The requirements of orthogonality in instruction encoding, and direct access to a large opcode space and programmer-visible register set, limit the size of immediate operands for arithmetic instructions to 8 bits. Only the LLI, LHI and LOLI instructions allow the encoding of halfword-sized immediates (16-bits). Table~\ref{Tab:byorisc:isa} summarizes the instruction set.

The instruction set is subdivided into instruction groups for arithmetic (A), load/store (LS), multiply (M), division (D), logical (L), set/comparison (C), immediate constant load (I), type conversion (T), control-transfer (F), procedure call (P) and coprocessor access (CP). The custom instruction group is denoted as CI. 

\begin{table}
  \centering
  \caption{The {ByoRISC} predefined instruction set.} 
  \begin{tabular}{|p{0.125\textwidth}|l|p{0.275\textwidth}|}
    \hline
    Mnemonic & Type & Description \\
    \hline
    LLI, LHI, LOLI & I & Halfword load (lower/upper/lower with OR)\\ 
    \hline
    LW, LB, LBU, LH, LHU, SW, SB, SH & LS & Load/store signed/unsigned words, bytes, and halfwords to/from a register\\ 
    \hline
    ADD, ADDU, SUB, SUBU & A & Arithmetic operation on two registers (rs, rt) and result write-back to rd\\ 
    \hline
    CVT & T & Convert type and transfer register\\ 
    \hline
    MUL, MULU & M & Multiply\\ 
    \hline
    DIV, DIVU & D & Divide\\ 
    \hline
    AND, OR, XOR, NOR & L & Logical operation on two registers (rs, rt) and write result to register rd\\ 
    \hline
    SRAV, SRLV, SLLV & S & Shift by a register operand\\ 
    \hline
    SRA, SRL, SLL & S & Shift by an immediate value\\ 
    \hline
    SEQ, SNE, SLT, SLTU, SLE, SLEU & C & Compare registers (rs, rt) and set register rd on condition\\ 
    \hline
    J, JR & F & Direct/indirect unconditional jump\\ 
    \hline
    BNEZ, BEQZ & F & Conditional branch\\ 
    \hline
    JAL & P & Jump and link to address (procedure call)\\ 
    \hline
    [C\textbar{}M][F\textbar{}T]CX, [L\textbar{}S]WCX & CP & Coprocessor interface instructions\\
    \hline
    SYSCAL, BREAK, HALT & -- & Software interrupt instructions\\ 
    \hline
  \end{tabular}
  \label{Tab:byorisc:isa}
\end{table}

A minimal ByoRISC has to support the following 22 instructions directly in hardware: add, addu, sub, subu, and, or, xor, lw, sw, lli, lhi, loli, srav, srlv, sllv, slt, sltu, j, jr, bnez, beqz, halt. 

\subsection{Custom instruction support in {ByoRISC} processors}
\label{subsec:byorisc:ci}

\subsubsection{Decoding of custom instructions}
\label{subsubsec:byorisc:cidec}

\begin{figure}[htb]
  \centering
  \includegraphics[width=8.0cm]{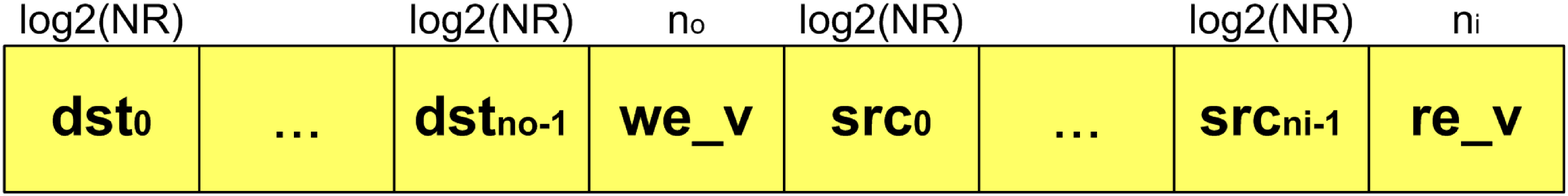}
  \caption{The SID lookup table entry format.}
  \label{Fig:byorisc:sid}
\end{figure}

For decoding CIs, the concept of Se\-condary Instruction Decoding (SID) has been introduced. SID is 
a variant of the concept of environment substitution \cite{Smotherman10}, used in the 
interpretation of microinstructions. SID applies this technique to program macroinstructions 
using programmer-invisible registers for the predecoding of CI operand addresses.
SID operation takes place in a partial decoding stage preceding the actual ID stage for base instructions, 
where CIs are identified based on their opcode MSBs. In the SID stage resides a lookup table (LUT) where 
the input and output register operand addresses for specific CI occurrences in user programs are kept. 
The LUT is addressed by the {\it ciocc} field of B-fmt instructions. An entry in the SID LUT is 
partitioned as shown in Fig.~\ref{Fig:byorisc:sid}, where: 

\begin{itemize}
  \item{$NR$ is the number of registers in the integer register file}
  \item{$n_{i}$, $n_{o}$ is the maximum allowable number of input and output operands of a CI}
  \item{$dst_{0}$ \ldots $dst_{n_{o}-1}$ and $src_{0}$ \ldots $src_{n_{i}-1}$ is the register address for output and input operands, respectively}
  \item{$we\_v$ and $re\_v$ is the write/read enable vector for output/input operands, correspondingly.}
\end{itemize}

For a requirement of $n_{i}$ input and $n_{o}$ output register operands, SID LUT entries have a width of: $(n_{i}+n_{o})\cdot(\log_{2}(NR)+1)$ which is simplified to: $2\cdot n \cdot(\log_{2}(NR)+1)$ given that $n_{i}=n_{o}=n$. For $n=8$ and $NR=256$, which is the typical case of a {ByoRISC} testbed architecture, each entry has a width of 144 bits. This implies allocating only 2 block RAMs in the Spartan-3 FPGA technology process (18k storage bits) for realizing a 256-entry 144-bit wide LUT, in a special, single read port, block RAM configuration mode. 

\subsubsection{Accessing operands of MIMO custom instructions}
\label{subsubsec:byorisc:ciopds}

The operand interface between the ByoRISC register resources and the datapath has to provide sufficient bandwidth in order not to compromise the performance benefits of tightly-coupled ASHEs. In this context, recent approaches such as \cite{Saghir07}, involve the utilization of a multi-port register file (MPRF) for zero-cycle overhead access to registers. Possible topologies of an MPRF involve:

\begin{itemize}
  \item A monolithic register file using a single register bank with multiplexer networks for direct read and write access to all registers. The required interconnection networks include full crossbar switches or permutation networks. Such a solution has been proved feasible for ASIC processes, but in contemporary fine-grain FPGAs, deep multiplexer staging is problematic in terms of both performance and dedicated resources.
  \item A clustered register file with block storage for single register copies. Generally, $n_{i}$ blocks of memory (assuming $n_{i}>n_{o}$) are required with one read and one write port each, allocating a single entry for each live variable. This topology has the advantage of not requiring multiplexing for register address decoding. However, simultaneous access to any register is not sustained, e.g. in the case that two registers are mapped to the same physical block resource. The register allocation algorithm should minimize such conflicts by creating multiple copies for these variables.
  \item A third solution is the use of a register file comprised of $n_{i} \times n_{o}$ block RAMs for maintaining the maximum number of multiple copies \cite{Saghir07}. In this topology, the zero overhead access to register operands is ensured. However, the demand in block storage resources may prove overwhelming especially in small FPGA devices. Since each memory bank is used for storing only $NR/n_{o}$ registers, the block RAMs are underutilized.
\end{itemize} 

The MPRF generator, {\it mprfgen}, that is used in the ByoRISC toolchain is freely available in source code form\footnote{\url{http://www.nkavvadias.com/misc/mprfgen.zip}}. {\it mprfgen} follows the approach by \cite{Saghir07} and has been successfully tested with up to Xilinx XST 12.3. Another MPRF generator has been recently reported in \cite{LaForest10}. 

\subsubsection{Data memory accesses}
\label{subsubsec:byorisc:cpar}

In {ByoRISC} processors, tightly-coupled ASHEs can have direct access to the data memory 
as regular operations in the context of a local FSMD (Finite-State Machine with Datapath). 
This capability allows for incorporating an arbitrary number and combination of load/store 
operations within CIs, however 
without eliminating the imposition of pipeline bubbles at all. Only a single 
data memory transfer is allowed at each processor cycle either it comprises a base instruction or 
part of the active CI at the time. 

\subsection{The microarchitecture of {ByoRISC} processors}
\label{subsec:byorisc:uarch}

\subsubsection{The configuration space of {ByoRISC} processors}
\label{subsubsec:byorisc:cfg}

A prominent characteristic of the {ByoRISC} architecture is the multi-parametric space, composed of more than 20 parameters, that is used for user-defined configuration of the microarchitecture description prior logic synthesis. The parameter set is given in Table~\ref{Tab:byorisc:cfg}.

\begin{table}[htb]
  \centering
  \caption{The configuration space of the {ByoRISC} microarchitecture.} 
  \begin{tabular}{|p{0.1\textwidth}|p{0.3375\textwidth}|}
    \hline
    Parameter & Description \\
    \hline
    HAVE\_* & Support for custom instructions (CI), local coprocessors (COP), 
              zero-overhead loop control (ZOLC), 8-bit immediates (SMALL\_IMM)\\ 
    \hline
    FORWARDING & Full data forwarding via scalable register bypassing\\
    \hline
    BR\_EARLY & Control flow transfer prior or after the EX/MEM pipeline register\\
    \hline
    [I\textbar{}D]MEMSIZE & Program/data memory size\\
    \hline
    OW, RAW & Opcode/register address width\\
    \hline
    NWP, NRP & Number of register file write/read ports\\
    \hline
    OPT\_* & Enabling optional features as: load/store instructions for small data types (LS), 
             shift-by-immediate instructions (SHIFT), 
             additional control-transfer instructions (CTI), type conversions (CVT), 
             hardware multiplier (MUL), divider (DIV), additional comparison (SET) 
             and logical instructions (LOGIC)\\
    \hline
    MULT\_TPL & Multiplier topology (single-cycle or 4-cycle latency pipelined)\\
    \hline
    SHIFTER\_TPL & Shifter topology (funnel, barrel, dedicated shifters)\\
    \hline
  \end{tabular}
  \label{Tab:byorisc:cfg}
\end{table}

\subsubsection{Integration of the ZOLC architecture}
\label{subsubsec:byorisc:zolc}

The ZOLC is a zero-overhead loop controller \cite{Kavvadias08} supporting arbitrary loop structures with multiple-entry and multiple-exit nodes that can be integrated in the instruction fetch ({IF}) stage of embedded RISC processors. A control flow graph (CFG) representation is used for procedures of the targeted application, for which the notion of control transfer expressions among CFG regions defined at loop boundaries is used. The instruction lists that comprise each of these regions are the {\it Data-Processing Tasks} (DPTs) of the CFG. The DPTs are distinguished between backward tasks that are involved in task switching decisions and update the loop index context of the CFG and forward tasks that may be involved in task switching but do not affect loop indices.

For ZOLC operation, the machine instructions that are involved in looping (loop index update, comparison to boundary values, and branching to the entry program counter -- PC -- of the succeeding DPT) are eliminated. Instead, the necessary task switching takes place during the {IF} of the last useful instruction of the specific DPT. Thus, no machine instructions are required for controlling the operation of ZOLC. The purpose of ZOLC is to provide a proper candidate PC target address to the PC decoding unit for each substituted looping operation.

\subsubsection{Pipeline organization}
\label{subsubsec:byorisc:pipeorg}

\begin{figure*}[htb]
  \centering
  \includegraphics[width=0.75\textwidth]{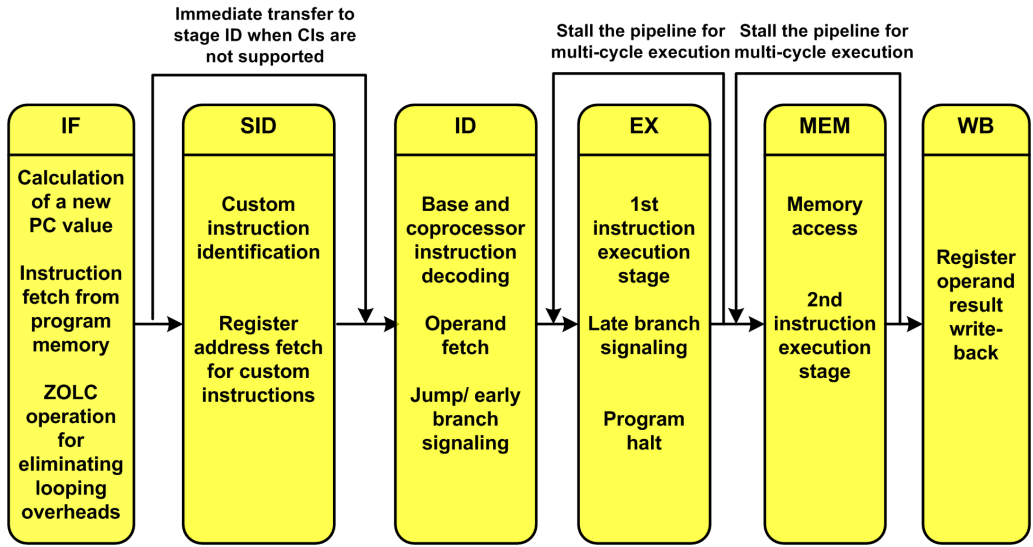}
  \caption{Pipeline organization for {ByoRISC}.}
  \label{Fig:byorisc:pipeorg}
\end{figure*}

A microarchitecture for {ByoRISC} has been fully implemented with a configurable 5/6 stage pipeline as shown in Fig.~\ref{Fig:byorisc:pipeorg}.

This organization is based on the classic 5-stage pipeline design encountered in popular embedded RISCs. The primary difference compared to traditional embedded processors is the addition of an intermediate pipeline stage, SID, succeeding instruction fetch and preceding the main instruction decoding stage. The execution stages (EX, MEM) can perform multi-cycle operations, while the hardware automatically stalls the preceding pipeline stages with the use of a stall clock. The storage resources (instruction and data memory, register file, SID LUT) have been designed for synchronous read operation for mapping to FPGA block RAMs by code inference.   

At the IF stage, an instruction is fetched from the program memory. In case ZOLC hardware is present and operates in active state, the corresponding PC value is chosen based on the DPT switching decision of ZOLC to the proper task entry PC of the subsequent DPT. Decoding of CI register operands is performed in the SID interim stage as described in section~\ref{subsubsec:byorisc:cidec}. The decoding of base and coprocessor instructions and the operand fetch for all instructions take place at the ID stage.

Stage EX is the first execution stage accessible to CIs. It is also used for datapath computations of base instructions. As operands, either the values fetched during ID or the forwarded ones from the full-bypassing data forwarding network can be used. The primitive base instructions are serviced by the ALU, the variable shifter and the branch unit. The basic addressing mode for these instructions is register direct, while additional modes can be introduced as user-defined extensions. An optional single- or four-cycle multiplier and a radix-2 divider are added when the corresponding derived instructions should be available in hardware. The four-cycle version of the multiplier could use 3 18$\times$18-bit embedded multipliers in Xilinx FPGAs (Spartan-3/3E, Virtex-4) when implemented with EX subpipeline stages for improved throughput. The EX stage also supports multi-cycle operation and is assigned with the task of communicating with the local coprocessors for transferring the necessary data. Coprocessor units are interfaced through a point-to-multipoint bus and cannot be scheduled in parallel with the core functional units; i.e. they are mutually exclusive to them and thus the pipeline need be stalled.

At the MEM stage, load and store base instructions access the data memory. In addition, CIs may interface to the data memory when operating in LOAD, STORE or SPECIAL\_CS (CI access) computational states. The final pipeline stage, WB, is responsible for committing destination register operands to the centralized register file as those are calculated by base, coprocessor and custom instructions.
  
\subsubsection{Scalable register bypassing (SRB) scheme}
\label{subsubsec:byorisc:srb}

A scalable scheme for full register bypassing in ByoRISC processors has also been developed \cite{Kavvadias09}. The register bypassing specification is parameterized regarding the number of homogeneous register file read/write ports and the number of execution pipeline stages of the processor. An abstract view of the proposed register bypassing scheme assumes a processor with a pipeline organization incorporating: 

\begin{itemize} 
\item{an instruction decode and operand fetch stage for reading $N_{RP}$ register operands}
\item{$N_{PIPE}$ execution stages with at least one of them accessing the data memory (for a typical ByoRISC it is: $N_{PIPE} = 2$)}
\item{a register write-back stage for writing $N_{WP}$ register operands}
\end{itemize}

The basic assumption for the first execution stage (EX1) is that it receives up to $N_{RP}$ read register operands from an MPRF and produces a result vector of up to $N_{WP}$ write register operands. The subsequent execution stages accept the result vector from their preceding stage, which is of width $N_{WP} \times DW$, where $DW$ is the register word width. Further, it can be specified that they read up to $N_{RP}$ from the forwarded read operands, given that these have been stored in the pipeline registers of the previous stage. The final pipeline stage is responsible for committing the final result vector to the register file. Any of the $N_{PIPE}$ execution stages can be configured for multi-cycle execution, stalling the previous ones for the required number of cycles.

The bypass network produces the multiplexer control signals that are used within EX1 for forwarding the appropriate data value. EX1 incorporates a set of multiplexers for selecting one of the forwarded values per register file read port.

The SRB hardware mainly comprises of the following components:

\begin{itemize} 
\item{$N_{RP} (N_{PIPE} \times N_{WP} + 1)$-to-1 multiplexers in EX1 for selecting the proper forwarded datum per read port.}
\item{$N_{RP} \times N_{PIPE} \times N_{WP}$ comparators for evaluating the multiplexer control signals. In case of supporting multi-cycle execution, the result of each comparator is AND-gated with a flag stating the completion of multi-cycle operation for the corresponding pipeline stage.}
\end{itemize}

Each of the EX1 multiplexers requires a control signal of width 
$\lceil log_{2}(N_{WP}) \rceil + \lceil log_{2}(N_{PIPE}+1) \rceil$. The multiplexer control signal format can be subdivided into two fields: field `pipe\_sel' which selects the appropriate pipeline execution stage for obtaining an intermediate result, with 0-th order referring to the register operand read stage and field `wp\_sel' for denoting a specific write port enumeration.

\begin{figure}[htb]
  \centering
  \includegraphics[width=0.4375\textwidth]{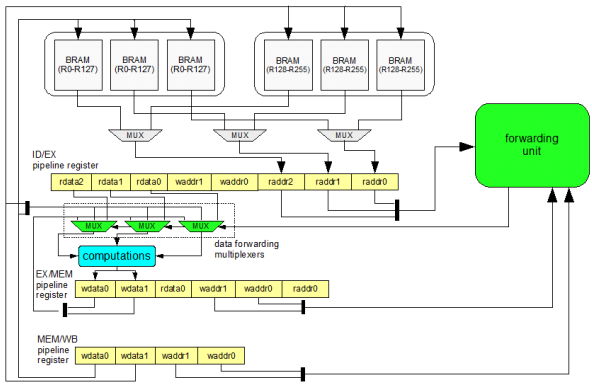}
  \caption{Scalable register bypassing for a ByoRISC processor instance.}
  \label{Fig:brisc:fwd}
\end{figure}

A detailed partial view of a 6-stage pipeline ByoRISC architecture is shown in Fig.~\ref{Fig:brisc:fwd}. In the figure, the bypass network (forwarding unit) and the data forwarding multiplexers as well as their associated interconnections can be easily identified. The MPRF has 3 read ports and 2 write ports and is implemented by 6 embedded memory blocks. The pipeline stage registers are used to appropriately pass the read data vector ($rdata0$ to $rdata2$), the read operand addresses ($raddr0$ to $raddr2$), and the write operand addresses ($waddr0$ to $waddr1$). The write data vector ($wdata0$ to $wdata1$) is propagated accordingly following its generation at the EX$n$ stage of the processor pipeline.

\subsection{The YARDstick custom instruction generation tool}
\label{sec:dse}

YARDstick is a CI generation and selection prototype framework.
Its main role is to facilitate design space exploration in heterogeneous flows for ASIP design 
where the development tools (compiler, binary utilities, simulator/ debugger) lack such capabilities. 

\begin{figure*}[htb]
  \centering
  \includegraphics[width=0.675\textwidth]{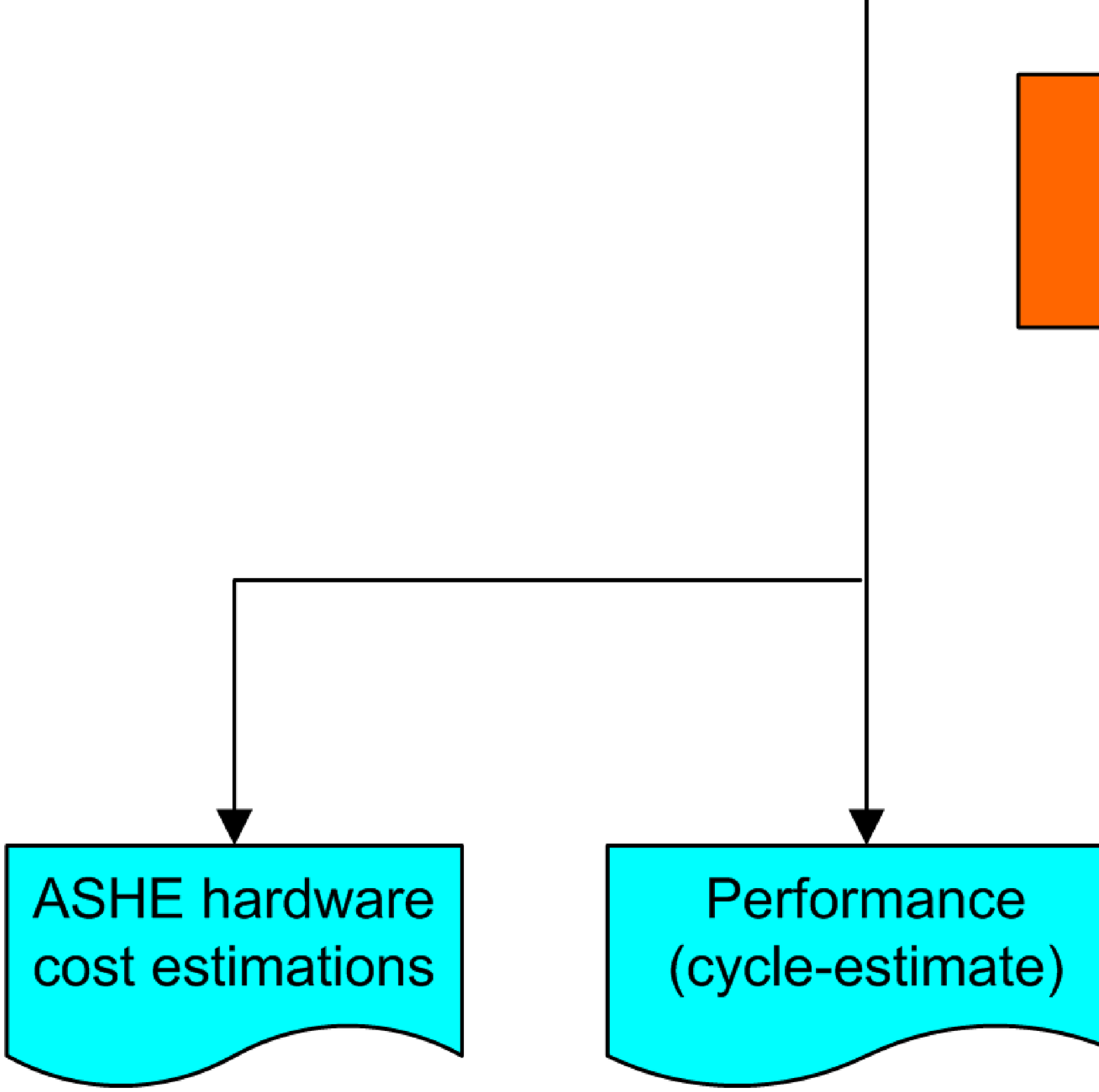}
  \caption{The YARDstick prototype framework.}
  \label{Fig:byorisc:yard:flow}
\end{figure*}

YARDstick is illustrated in Fig.~\ref{Fig:byorisc:yard:flow}. It accepts input in ANSI C through 
the SUIF2 frontend,\footnote{\url{http://suif.stanford.edu/suif/suif2/index.html}} subsequently lowered 
to Machine-SUIF IR \cite{MachSUIF}, assembly code or directly to a textual IR 
in the form of a flat CDFG. The latter form is termed as `ISeq' 
(Instruction Sequence). The resulting IR can use {\it SUIFvm}, 
{\it SUIFrm} (introducing physical registers to {\it SUIFvm}) or machine-specific 
({ARMv4}, {DLX}, {ByoRISC} have been tested) instruction semantics. For {\it SUIFvm} and {\it SUIFrm}, 
complete procedure entry and exit sequences are not inserted at this stage, since 
stack frame layout is highly processor dependent. 

In the following stage, Machine-SUIF passes are used for performing analyses (static instruction mix, 
data type analysis) and classic compiler scalar optimizations. A peephole matching-based code 
selection pass is then applied. 
The resulting assembly-level code can then be macro-expanded, instrumented for profiling and converted to {ISeq} 
by an appropriate SALTO pass \cite{SALTO}. For each target architecture, a working SALTO backend library 
must have been developed. Assembly code can be processed by the target 
machine binary utilities (auto-generated \ien{binutils} port from the corresponding ArchC model) 
and the resulting ELF executables can be run on an instruction- or cycle-accurate simulator. 
Alternatively, ISeq files can be generated as compiler IR dumps directly from the compiler for the 
target machine. This is the case for a modified version of Machine-SUIF \cite{MachSUIF} for 
which the basic block profile is automatically obtained by converting the IR to a C subset with the 
{\it m2c} pass and executing the low-level C code on the native machine.

The CI generation process takes place on the optimized IR as well, and is 
then followed by CI selection. In order to drive CI generation, 
the target specification is given in the so-called BXIR (Build your own Compiler-Simulator IR) form along with 
the dynamic profile of the application. BXIR entries 
contain information on the inputs/outputs, area demand, fractional latency and 
required cycles of each hardware operator. In general, each IR-level operation is 
assumed to be implemented by a dedicated hardware operator. Both latency and area 
metrics are scaled against the dominant operator in the given BXIR specification, 
which usually is the hardware multiplier or divider.

A number of CI generation methods have been implemented involving the identification of
MIMO or MISO (Multiple-Input Single-Output) CIs under user-defined constraints. These methods are:
\begin{itemize}
\item {MAXMISO \cite{Alippi99} for identifying maximal subgraphs with a single-output node using a linear complexity algorithm.}
\item {MISO exploration under constraints \cite{Kavvadias05}.} 
\item {MIMO CI generation \cite{Pozzi06}. As a pruning policy, a fast heuristic is employed by assuming similarly to \cite{Pothineni07}, that the performance gain provided by a pattern $P$ is higher than that of any subgraph of $P$\footnote{The monotonicity property of the convex subgraph speedup model was proved formally in \cite{Verma07}}. The user can disable this option and apply the exponential complexity algorithm, e.g. if all valid subgraphs must be enumerated.}
\end{itemize}

Regarding CI selection, an optimal 0-1 knapsack-based and a greedy method based on
predefined priority metrics have been implemented. Graph isomorphism is used to identify the unique CI 
patterns, while applying graph-subgraph isomorphism is used for identifying the 
patterns corresponding to unique extension units, servicing a subset of generated instructions. 
The matching process can take account individual opcodes or resource classes. Different 
instructions with opcodes of the same class can be matched and considered to be 
implemented on the same basic resource. The used graph isomorphism
algorithms are part of the VFLib2 graph matching library 
\cite{Foggia01}. 

As outcomes of using YARDstick on an input application, estimated hardware costs
for the extension units and a cycle estimate for the given application are obtained. A prototype code 
generator ({\it codesel}) is used for mapping the generated CIs on the given application.
Then, the cycle-estimate ArchC model for the architecture is automatically linked to the 
C models of the selected CIs.

YARDstick incorporates a number of backend engines for the generation of:
\begin{itemize}
\item {ANSI C subset code for incorporation to user tools (ArchC simulators, validators, behavioral synthesizers).
This code is used as input to the SPARK \cite{SPARK} high-level synthesis tool for generating 
the RTL description of the CI hardware.}
\item {GDL/VCG \cite{VCG} and Graphviz \cite{Graphviz} files for visualization of application 
call graphs, control-flow graphs, basic blocks and CIs.}
\item {An extended CDFG \cite{CDFGtool} format for scheduling and translation to dataflow VHDL.}
\item {Export to an XML format which is supported by the AGG \cite{AGG} attributed graph 
transformation system.}
\end{itemize}

\section{Performance/area characterization of a representative {ByoRISC}}
\label{sec:perf}
In order to evaluate the timing (critical path) and area of a 
typical {ByoRISC} processor, a configured instance of the parameterized VHDL model of 
the processor was used. Directives for the $vpp$ VHDL preprocessor \footnote{\url{http://sourceforge.net/projects/vpp}}
are used in order to parameterize the actual VHDL model. The testbench code uses the PCK\_FIO package for 
{\it printf}-style output in VHDL\footnote{\url{http://www.easics.com/webtools/freesics}}. 

The complete VHDL model of ByoRISC amounts about 6.5k lines of code (LOC) that can be subdivided into the 
four classes depicted in Table~\ref{Tab:brisc:sloc}. This model assumes the existence of only a dummy CI. 
However, a library of CI implementations such as for alpha blending, 
population count, clipping, counting leading zeroes/ones has been manually designed and tested by the 
author. Further, a complete file listing of ByoRISC (with a sample CI) is also available from the 
author's website.

\begin{table}
  \centering
  \caption{Lines-of-code for the VHDL model of ByoRISC.} 
  \begin{tabular}{|l|p{0.3125\textwidth}|p{0.0375\textwidth}|}  
    \hline
    Class & Description & LOC \\
    \hline
    A & VHDL RTL code passed through vpp (with pp\_ prefix) & 3631 \\
    \hline
    B & VHDL RTL code that doesn't use vpp & 1376 \\
    \hline
    C & VHDL RTL code for the sample system & 758 \\
    \hline
    D & Testbench code (non-synthesizable) including PCK\_FIO & 754 \\
    \hline
      & Total & 6519 \\
    \hline
  \end{tabular}
  \label{Tab:brisc:sloc}
\end{table}

The ByoRISC core for the following experiments is an out-of-the-box configuration
including instruction block RAM memory, excluding data memory (the latter is instantiated 
as part of ByoRISC systems), while supporting CIs, 
all additional instructions except division, type conversion and auxiliary comparisons. A pipelined
4-cycle multiplier and a funnel shifter are also instantiated. A size of 8KB 
is used for the separate cacheless instruction and data memories. The model supports up to 256 primary opcodes, 
8 input and 8 output operands for each CI and 256 physical registers. It
only contains a skeleton CI unit that can be configured to perform a permutation of 
8 inputs to 8 outputs by plain wiring.

For each case, the timing and area requirements are estimated with the help of the Mentor LeonardoSpectrum (ASIC) and Xilinx Webpack ISE 7.1.04i (FPGA) synthesis tools. One of the smallest Virtex-4 Xilinx FPGAs was selected, namely the XC4VLX25 device (`-10' speed grade), which incorporates 21504 LUTs, 72 18-kbit block RAMs (BRAMs) and 48 DSP48 embedded datapaths.

\begin{figure}[htb]
  \centering
  \subfigure[Standard cell VLSI STM 0.13$\mu$m.]{
  \framebox{\includegraphics[width=0.4375\textwidth]{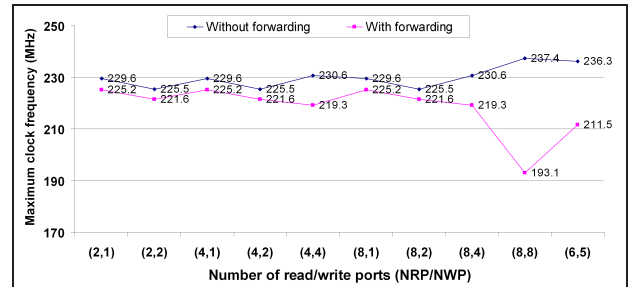}}
  \label{Fig:perf:brisc:delay:asic}}
  \subfigure[Virtex-4 device: XC4VLX25]{
  \framebox{\includegraphics[width=0.4375\textwidth]{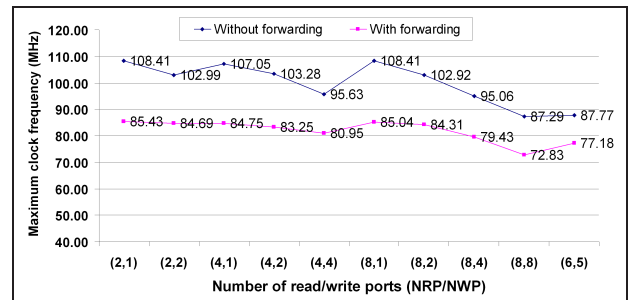}}
  \label{Fig:perf:brisc:delay:fpga}}
  \caption{Maximum clock frequency for the {ByoRISC} processor.}
  \label{Fig:perf:brisc:delay}
\end{figure}

\begin{figure}[htb]
  \centering
  \subfigure[Standard cell VLSI STM 0.13$\mu$m.]{
  \framebox{\includegraphics[width=0.4375\textwidth]{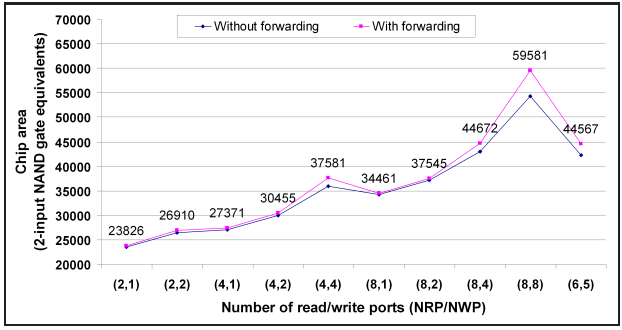}}
  \label{Fig:perf:brisc:area:asic}}
  \subfigure[Virtex-4 device: XC4VLX25]{
  \framebox{\includegraphics[width=0.4375\textwidth]{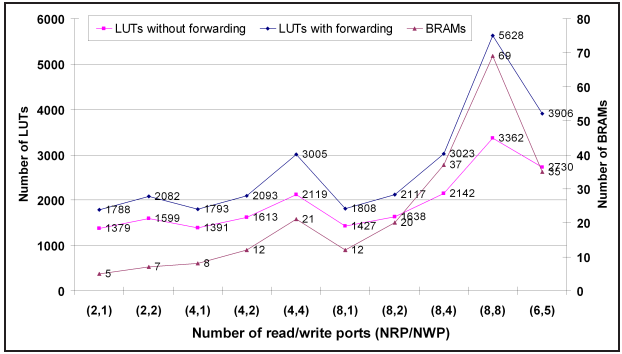}}
  \label{Fig:perf:brisc:area:fpga}}
  \caption{Chip area for the {ByoRISC} processor.}
  \label{Fig:perf:brisc:area}
\end{figure}

Fig.~\ref{Fig:perf:brisc:delay} depicts the maximum clock frequency estimates for different number of supported read ($N_{RP}$) and write ($N_{WP}$) register file ports. The chip area requirements are shown for both processes in Fig.~\ref{Fig:perf:brisc:area}. The number of execution pipeline stages has been set to 2, since the automatic pipelining of CIs over multiple execution stages has not been considered.

A base ByoRISC with no forwarding requires 1379 LUTs, 5 BRAMs, and 3 DSP48 blocks. The latter remains unchanged for all configurations so it is not shown in Fig.~\ref{Fig:perf:brisc:area:fpga}. From these figures, it can be seen that the number of read/write ports escalates the chip area on the FPGA device to about 4 times in terms of LUTs, from 1379 to 5628, that is up to 26.4\% of the total LUT resources. ByoRISCs supporting CIs with many inputs and outputs have high demands on BRAMs. For $(N_i,N_o)=(8,8)$, 69 out of the 72 available BRAMs are required for the base ByoRISC. For the ASIC process, without accounting for the register file area, the corresponding value range is about three times compared to the baseline case figures (18k to 60k gates). In addition, the use of a full data forwarding network decreases the maximum clock frequency by 17.9\%. On the contrary, for the ASIC process, this performance degradation measures to only 5\%. The difference in maximum clock frequency among the (2,1) and (8,8) configurations with and without the use of full data forwarding, measures to 19\% and 9.7\%, respectively for the FPGA.

\section{Case study: An Image Processing Pipeline}
\label{sec:ipp}
\begin{figure}[htb]
  \centering
  \framebox{\includegraphics[width=8.0cm]{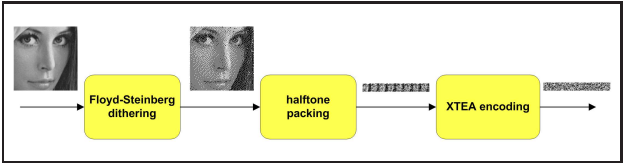}}
  \caption{An image processing flow comprising of dithering, data packing and XTEA encryption.}
  \label{Fig:ipp:flow}
\end{figure}

In order to evaluate the performance of the {ByoRISC} architecture on realistic applications, an image processing pipeline (IPP) has been used. The encoding flow of the IPP which is shown in Fig.~\ref{Fig:ipp:flow} processing 256-level greyscale images, comprises of three application kernels: {\it fsdither} (Floyd-Steinberg dithering by error diffusion to a bilevel image), {\it htpack} (halftone image packer for 8-fold lossless compression of a bilevel image), and {\it xteaenc} (XTEA encryption). A complementary pipeline for data decompression involves application 
kernels {\it htunpack} (halftone image unpacker) and the XTEA decoder ({\it xteadec}). An $n$-order Hilbert curve generator, which is an application not used in the IPP is also evaluated.

First, the critical basic blocks of the applications have been identified (Table~\ref{Tab:ipp:hotbbs}). It can be seen that these blocks comprise of the 99.5\% (almost totality) of the IPP encoding flow dynamic instruction cycles, so it is sufficient that generated CIs only account for these. For the performance-critical basic blocks, early measures of speedup potential were obtained. This was possible by computing an ASAP schedule with unlimited resources, by static analysis at the ISeq level using an enhanced version of the {\it asapalap} tool, part of an extended version of CDFGtool. For each time-critical basic block the following metrics were measured:
\begin{itemize}
\item {\it max\_ilp}: the maximum parallelism for a given control step in the schedule
\item {\it csteps}: the number of control steps for performing the schedule
\item {\it avg\_ilp}: the average operation-level parallelism, calculated as $num\_ops/csteps$, where $num\_ops$
is the number of operations in the corresponding basic block.
\end{itemize}

\begin{figure}[htb]
  \centering
  \framebox{\includegraphics[width=0.4375\textwidth]{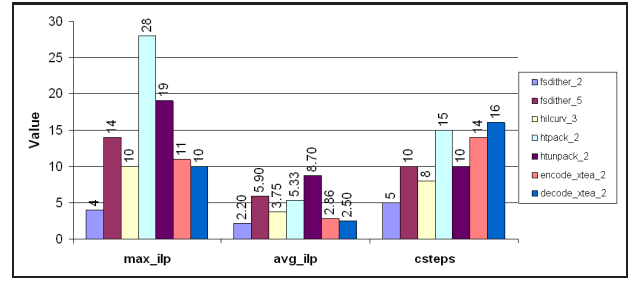}}
  \caption{Metrics of the available intrinsic parallelism for hot basic blocks 
  of IPP applications.}
  \label{Fig:ipp:ilp}
\end{figure}

Fig.~\ref{Fig:ipp:ilp} gives in detail the three quantities as calculated for the critical basic blocks of the IPP applications.
It is observed that with the exception of {\it fsdither}, the maximum useful parallelism is above 10, indicating significant performance potential for MIMO CI generation.
 
\begin{table}
  \centering
  \caption{Time-critical basic blocks for some IPP applications.} 
  \begin{tabular}{|l|c|c|c|}
    \hline
    Application & Instructions & Est. dynamic cycles & \% of application \\
    \hline
    fsdither.5 & 59 & 237568 & 85.19\\ 
    \hline
    fsdither.2 & 10 & 40960 & 14.69\\ 
    \hline
    htpack.2 & 80 & 40448 & 99.98\\ 
    \hline
    xteaenc.2 & 44 & 638976 & 95.63\\ 
    \hline
  \end{tabular}
  \label{Tab:ipp:hotbbs}
\end{table}

A greedy selector for the `cycle gain' priority metric has been used. A summary of the identified CIs is given in Table~\ref{Tab:ipp:cis}. The first column denotes the identified CI with the second providing the number of input $(N_i)$, output $(N_o)$, and constant $(N_c)$ operands used by the CI. 
The following columns provide measurements for cycle gain, incremental speedup (compared to a base case with no CIs), software cycles (cycles required for a sequential implementation of the CI on ByoRISC), hardware cycles (cycles required for executing the CI), and estimated area, respectively. These CIs have been identified by YARDstick using a $(N_i,N_o)=(8,8)$ constraint. The significant reduction in terms of hardware cycles is due to the strong effect of operation chaining, i.e. grouping of several dependent operations within the same cycle, which is evident for constant shift, logical and bit manipulation operations. A speedup improvement of 4.9$\times$ is expected over the application set based on the estimations (averaging the results of column `Incr. speedup' over the applications of IPP) made by YARDstick.

\begin{table}
  \centering
  \caption{CI characteristics.} 
  \begin{tabular}{|l|l|p{0.05\textwidth}|p{0.05\textwidth}|p{0.03\textwidth}|p{0.03\textwidth}|p{0.04\textwidth}|}
    \hline
    CI & $N_{i}$,$N_{o}$,$N_{c}$ & Cyc. gain & Incr. speedup & SW cyc. & HW cyc. & Area (MAU) \\
    \hline
    fsdither0 & 5,2,9 & 180224 & 2.69 & 57 & 13 & 4.56\\
    \hline
    fsdither1 & 3,2,2 & 32768 & 3.87 & 9 & 1 & 0.18\\
    \hline
    hilcurv0 & 5,5,8 & 638976 & 5.46 & 28 & 2 & 0.35\\
    \hline
    htpack0 & 3,2,14 & 35328 & 7.25 & 78 & 9 & 1.10\\
    \hline
    htunpack0 & 5,2,12 & 36864 & 5.79 & 81 & 9 & 8.24\\
    \hline
    xteaenc0 & 6,5,7 & 540672 & 4.55 & 38 & 5 & 0.83\\
    \hline
    xteadec0 & 6,5,7 & 540672 & 4.70 & 38 & 5 & 0.78\\
    \hline
  \end{tabular}
  \label{Tab:ipp:cis}
\end{table}

The benchmark statistics for various scenarios are summarized in Table~\ref{Tab:ipp:summary}.
Columns 2--7 provide measurements on the given application set. The last column illustrates 
the weighted average for corresponding estimates of application metrics.
Line ``Initial cycles'' gives the dynamic execution cycles  
on ByoRISC without CIs. ``Cyc. with CIs'' refers to the same metric when CIs 
are enabled. ``App. speedup'' lines (5th and 6th) illustrate the actual speedup achieved 
in hardware and the estimate by YARDstick, respectively. When ZOLC is enabled, the 
dynamic cycles are further reduced as shown in line ``Cyc. with CIs-ZOLC''. The 
corresponding application speedup due to the ArchC simulation is 
given in the following line. Line ``\% diff.'' provides the percentage 
difference regarding lines 5--6.

Table~\ref{Tab:ipp:summary} shows that the actual speedup is about 4.4$\times$, 
meaning that the high-level estimations made by YARDstick have an error of about 12\%.
When the ZOLC is enabled, the weighted speedup is about 5.7$\times$, due to a further cycle 
reduction of about 30\% compared to using CIs without the effect of ZOLC. 
Such results are to be expected for small application kernels; previous 
work \cite{Kavvadias08} gives about 25\% speedup improvement for kernels and 
10\% for entire applications.

\begin{table*}
  \centering
  \caption{A summary of benchmark statistics.} 
  \begin{tabular}{|p{0.1375\textwidth}|p{0.05\textwidth}|p{0.05\textwidth}|p{0.05\textwidth}|p{0.05\textwidth}|p{0.05\textwidth}|p{0.05\textwidth}|p{0.0625\textwidth}|}
    \hline
    \multicolumn{1}{|m{0.1\textwidth}|}{\centering Description}
    &\multicolumn{6}{p{0.3\textwidth}|}{\centering Applications}
    &\multicolumn{1}{p{0.0625\textwidth}|}{\centering Overall}\\  
    \hline
     & fsdither & hilcurv & htpack & htunpack & xteaenc & xteadec & Weighted average \\
    \hline
    Initial cycles & 250248 & 725009 & 51421 & 57013 & 619033 & 588310 & \\ 
    \hline
    Cyc. with CIs & 74133 & 143377 & 7183 & 8207 & 158230 & 144406 & \\ 
    \hline
    App. speedup (VHDL sim.) & 3.38 & 5.06 & 7.16 & 6.95 & 3.91 & 4.07 & 4.41 \\
    \hline
    App. speedup (YARDstick est.) & 3.87 & 5.46 & 7.25 & 7.39 & 4.70 & 4.55 & 4.94 \\
    \hline
    Cyc. with CIs-ZOLC & 37269 & 114705 & 6671 & 6671 & 141334 & 110614 & \\
    \hline
    App. speedup (CIs-ZOLC, sim.) & 6.71 & 6.32 & 7.71 & 8.55 & 4.38 & 5.32 & 5.67 \\
    \hline
    \% diff. & 12.77 & 7.39 & 1.26 & 6.00 & 16.76 & 10.46 & 12.01 \\
    \hline
  \end{tabular}
  \label{Tab:ipp:summary}
\end{table*}

\begin{figure}[htb]
  \centering
  \includegraphics[width=0.4375\textwidth]{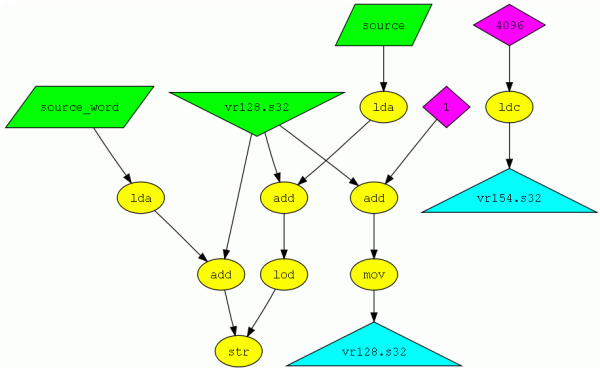}
  \caption{Sample CI (\ten{fsdither1}) generated by YARDstick from 
  application \ien{fsdither}.}
  \label{Fig:ipp:ises}
\end{figure}

Fig.~\ref{Fig:ipp:ises} illustrates the data-dependence graph of a sample CI from the 
image processing benchmark set, namely \ten{fsdither1}. 

\subsection{Performance comparison against VEX}
\label{sec:ipp:vex}
In another set of experiments, ByoRISC has been evaluated against 
a parameterized VLIW architecture named VEX\footnote{\url{http://www.hpl.hp.com/downloads/vex/}} 
also described in detail in \cite{Fisher04}. The VEX toolchain provides the means to 
target a wide class of embedded VLIW processors, by using a complete ANSI C compilation 
toolset and a cycle-accurate simulator. 
VEX was configured as a single-cluster VLIW machine featuring a configurable 
number of slots: 1, 2, 4, 8, and 16. The \ten{-h2 -O3} compilation options were used, 
enabling data-oriented optimizations such as aggressive loop 
unrolling. The VEX scheduler, attempts to 
schedule the maximum available number of independent operations in parallel, which is 
a different approach to CI optimization, since the latter focuses 
both on grouping independent operations (spatial independence) and chained data dependencies 
(temporal dependence). 

An open-source VEX implementation\footnote{\url{http://code.google.com/p/r-vex}},
namely $\rho$-VEX, which employs a 4-wide VLIW architecture is comparable to a ByoRISC with $(N_i,N_o)=(8,4)$
based on their register file configurations. $\rho$-VEX has been synthesized with Xilinx ISE for 
the same device as in Section~\ref{sec:perf}; a maximum clock frequency of 56MHz and 
an area demand of 19523 LUTs and 14 DSP48 datapaths was revealed. ByoRISC including 
the corresponding ASHEs achieves a clock frequency of 79MHz and an area requirement of 10565 LUTs (due to 
additional 7542 LUTs for the CIs), 3 DSP48 units 
and 37 block RAMs. Thus, ByoRISC uses about half the LUTs of a VEX4, and about 
half of the available BRAM resources, mainly for the multi-port register file, 
while $\rho$-VEX uses distributed LUT RAM for implementing its register file. Obviously,  
it can be safely assumed that the maximum clock frequency for $\rho$-VEX could be 
used for operating both processors.

\begin{figure}[htb]
  \centering
  \framebox{\includegraphics[width=8.0cm]{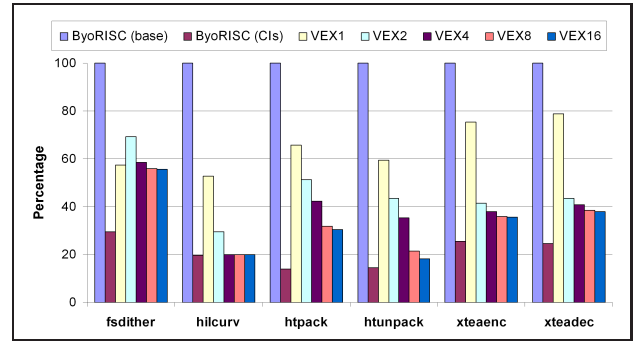}}
  \caption{Relative cycle counts for ByoRISC and VEX configurations.}
  \label{Fig:ipp:vex}
\end{figure}

Figure~\ref{Fig:ipp:vex} illustrates the relative cycle count for a base ByoRISC, 
a ByoRISC using the identified CIs and for the five different VEX configurations.
Table~\ref{Tab:ipp:vex} provides the exact cycles for VEX; corresponding counts 
for ByoRISC are already shown in Table~\ref{Tab:ipp:summary} (`Initial cycles' and `Cyc. with CIs', 
correspondingly.)

A first observation is that the initial cycles for ByoRISC are higher than of the 
RISC-like VEX configuration (VEX1) for all applications. This is due to the non-compact encodings used 
by ByoRISC; base ByoRISC instructions essentially comprise primitive operations without side-effects. 
However, the weighted speedup achieved by VEX when comparing its one-wide to the 16-wide configuration (VEX16) is 
about 2.23$\times$, about the half of the achieved speedup obtained by the CI concept on 
ByoRISC. ByoRISC outperforms VEX16 in five out of the six applications; 
VEX achieves slightly better results for the \ien{hilcurv} benchmark even with a four-wide configuration. 

\begin{table}
  \centering
  \caption{Absolute cycle counts for VEX configurations.} 
  \begin{tabular}{|p{0.1\textwidth}|r|r|r|r|r|}
    \hline
    Application & VEX1 & VEX2 & VEX4 & VEX8 & VEX16 \\
    \hline
     fsdither & 143768 & 99667 & 84053 & 80412 & 79949 \\
    \hline
     hilcurv  & 382248 & 214232 & 144607 & 144598 & 144596 \\
    \hline
     htpack   & 33871  & 26418 & 21811 & 16339 & 15621 \\
    \hline
     htunpack & 33947  & 24759 & 20152 & 12234 & 10445 \\
    \hline
     xteaenc  & 466824 & 256793 & 234399 & 222079 & 220167 \\
    \hline
     xteadec  & 463349 & 255375 & 240186 & 226287 & 223895 \\
    \hline
  \end{tabular}
  \label{Tab:ipp:vex}
\end{table}

\section{Conclusions}
\label{sec:sum}

In this paper, the configurable {ByoRISC} processor architecture has been presented. {ByoRISCs} are well suited to design space exploration due to their scalability; such an example being the multi-port register file and scalable data forwarding architecture. Further, {ByoRISC} processors allow the investigation of possibilities for ASHE integration. Hardware characterization of a reference {ByoRISC} model proves that this approach is feasible even on moderately sized FPGAs. A case study image processing application set was explored and implemented, unveiling a potential acceleration of 4.4$\times$ compared to the baseline processor. Further, {ByoRISC} outperforms a well-known academic VLIW architecture named VEX, in all tested applications except one, even when the VEX uses a 16-wide configuration.

DSE is enabled by YARDstick, which provides a compiler/ simulator-agnostic infrastructure for application analysis, performance estimation and CI generation. With YARDstick, the impact of register allocation, ASHE local storage and prioritized selection on the quality of CIs, generated under different input/output constraints, were investigated. For the aforementioned benchmark set, YARDstick provides an estimation within 12\% of the actual performance.


\bibliographystyle{IEEEtran}
\bibliography{IEEEabrv,byorisc-journal}

\begin{thebibliography}{10}
\providecommand{\url}[1]{#1}
\csname url@samestyle\endcsname
\providecommand{\newblock}{\relax}
\providecommand{\bibinfo}[2]{#2}
\providecommand{\BIBentrySTDinterwordspacing}{\spaceskip=0pt\relax}
\providecommand{\BIBentryALTinterwordstretchfactor}{4}
\providecommand{\BIBentryALTinterwordspacing}{\spaceskip=\fontdimen2\font plus
\BIBentryALTinterwordstretchfactor\fontdimen3\font minus
  \fontdimen4\font\relax}
\providecommand{\BIBforeignlanguage}[2]{{%
\expandafter\ifx\csname l@#1\endcsname\relax
\typeout{** WARNING: IEEEtran.bst: No hyphenation pattern has been}%
\typeout{** loaded for the language `#1'. Using the pattern for}%
\typeout{** the default language instead.}%
\else
\language=\csname l@#1\endcsname
\fi
#2}}
\providecommand{\BIBdecl}{\relax}
\BIBdecl

\bibitem{Gonzalez00}
R.~Gonzalez, ``Xtensa: A configurable and extensible processor,'' \emph{IEEE
  Micro}, vol.~20, no.~2, pp. 60--70, Mar.-Apr. 2000.

\bibitem{Sirowy07}
S.~Sirowy, Y.~Wu, S.~Lonardi, and F.~Vahid, ``Two-level microprocessor
  accelerator partitioning,'' in \emph{Proc. Design, Automation and Test in
  Europe Conf.}, Nice, France, Apr. 2007, pp. 313--318.

\bibitem{Goodwin03}
D.~Goodwin and D.~Petkov, ``Automatic generation of application specific
  processors,'' in \emph{Proc. Int. Conf. on Compilers, Architectures and
  Synthesis for Embedded Systems}, San Jose, California, USA, Oct. 2003, pp.
  137--147.

\bibitem{ClarkN05}
N.~Clark, J.~A. Blome, M.~L. Chu, S.~A. Mahlke, S.~Biles, and K.~Flautner, ``An
  architecture framework for transparent instruction set customization in
  embedded processors,'' in \emph{Proc. 32nd Int. Symp. on Computer
  Architecture}, Madison, Wisconsin, USA, Jun. 2005, pp. 272--283.

\bibitem{Halfill03}
T.~R. Halfill, ``{MIPS} embraces configurable technology,''
  \emph{Microprocessor {R}eport}, March 3 2003.

\bibitem{Leupers06}
R.~Leupers, K.~Karuri, S.~Kraemer, and M.~Pandey, ``A design flow for
  configurable embedded processors based on optimized instruction set extension
  synthesis,'' in \emph{Proc. Design, Automation and Test in Europe Conf.},
  Messe Munich, Germany, Mar. 2006.

\bibitem{Vassiliadis04}
S.~Vassiliadis, S.~Wong, G.~Gaydadjiev, K.~Bertels, G.~Kuzmanov, and
  E.~Panainte, ``The {MOLEN} polymorphic processor,'' \emph{IEEE Trans.
  Comput.}, vol.~53, no.~11, pp. 1363--1375, Nov. 2004.

\bibitem{Rosinger04}
H.-P. Rosinger, \emph{{XAPP529}: Connecting Customized {IP} to the {MicroBlaze}
  Soft Processor Using the {F}ast {S}implex {L}ink ({FSL}) Channel}, v1.3~ed.,
  May 2004.

\bibitem{Chen06}
X.~Chen and D.~Maskell, ``{M2E}: A multiple-input, multiple-output function
  extension for {RISC}-based extensible processors,'' in \emph{Architecture of
  Computing Systems -- ARCS 2006}, ser. Lecture Notes in Computer Science,
  W.~Grass, B.~Sick, and K.~Waldschmidt, Eds.\hskip 1em plus 0.5em minus
  0.4em\relax Springer Berlin / Heidelberg, 2006, vol. 3894, pp. 191--201.

\bibitem{Kavvadias08b}
N.~Kavvadias and S.~Nikolaidis, ``The {ByoRISC} configurable processor
  family,'' in \emph{Proceedings of the IFIP/IEEE VLSI-SoC 2008 --
  International Conference on Very Large Scale Integration}, Rhodes Island,
  Greece, October 13--15 2008, pp. 439--444.

\bibitem{Kavvadias07}
------, ``{YARDstick}: Automation tool for custom processor development,'' in
  \emph{presented at the University Booth of the Design, Automation and Test in
  Europe Conf.}, Nice, France, Apr. 2007.

\bibitem{SABRE}
S.~Chappell, A.~Macarthur, D.~Preston, D.~Olmstead, B.~Flint, and C.~Sullivan,
  ``Exploiting real-time {FPGA} based adaptive systems technology for real-time
  sensor fusion in next generation automotive safety systems,'' in \emph{In
  Proceedings of the Design, Automation and Test in Europe Conf.}, ser. DATE
  '05, Washington, DC, USA, 2005, pp. 180--185.

\bibitem{Kavvadias08}
N.~Kavvadias and S.~Nikolaidis, ``Elimination of overhead operations in complex
  loop structures for embedded microprocessors,'' \emph{IEEE Trans. Comput.},
  vol.~57, no.~2, pp. 200--214, Feb. 2008.

\bibitem{Campi01}
F.~Campi, R.~Canegallo, and R.~Guerrieri, ``{IP}-reusable 32-bit {VLIW} {RISC}
  core,'' in \emph{Proceedings of the 27th European Solid-State Circuits
  Conf.}, Villach, Austria, September 2001, pp. 456--459.

\bibitem{Smotherman10}
M.~Smotherman, ``A brief history of microprogramming,'' September 2010,
  http://www.cs.clemson.edu/~mark/uprog.html.

\bibitem{Saghir07}
M.~A.~R. Saghir and R.~Naous, ``A configurable multi-ported register file
  architecture for soft core processors,'' in \emph{Proc. 2007 Int. Workshop on
  Applied Reconfigurable Computing}, Mangaratiba, Rio de Janeiro, Brazil, Mar.
  2007, pp. 14--25.

\bibitem{LaForest10}
C.~E. LaForest and J.~G. Steffan, ``Efficient multi-ported memories for
  {FPGAs},'' in \emph{Proceedings of the 18th annual ACM/SIGDA international
  symposium on Field programmable gate arrays}, ser. FPGA '10, Monterey,
  California, USA, 2010, pp. 41--50.

\bibitem{Kavvadias09}
N.~Kavvadias and S.~Nikolaidis, ``Scalable register bypassing for {FPGA}-based
  processors,'' \emph{Microprocessors and Microsystems}, vol.~33, no. 7--8, pp.
  441--452, Oct.-Nov. 2009.

\bibitem{MachSUIF}
M.~D. Smith and G.~Holloway, ``Machine-{SUIF} research compiler,'' 2002,
  http://www.eecs.harvard.edu/hube/software/.

\bibitem{SALTO}
E.~Rohou, ``{SALTO}: {S}ystem for assembly-language transformation and
  optimization,'' 1997, http://www.irisa.fr/caps/projects/Salto/.

\bibitem{Alippi99}
C.~Alippi, W.~Fornaciari, L.~Pozzi, and M.~Sami, ``A {DAG} based design
  approach for reconfigurable {VLIW} processors,'' in \emph{Proc. Design,
  Automation and Test in Europe Conf.}, Munich, Germany, Mar. 1999, pp.
  778--779.

\bibitem{Kavvadias05}
N.~Kavvadias and S.~Nikolaidis, ``Automated instruction-set extension of
  embedded processors with application to {MPEG}-4 video encoding,'' in
  \emph{Proc. 16th Int. Conf. on Application-specific Systems, Architectures
  and Processors}, Samos, Greece, Jul. 2005, pp. 140--145.

\bibitem{Pozzi06}
L.~Pozzi, K.~Atasu, and P.~Ienne, ``Exact and approximate algorithms for the
  extension of embedded processor instruction sets,'' \emph{IEEE Trans. CAD
  Integr. Circ. Syst.}, vol.~25, no.~7, pp. 1209--1229, Jul. 2006.

\bibitem{Pothineni07}
N.~Pothineni, A.~Kumar, and K.~Paul, ``Application specific datapath extension
  with distributed {I/O} functional units,'' in \emph{Proc. 20th Int. Conf. on
  VLSI Design}, Bangalore, India, Jan. 2007, pp. 551--558.

\bibitem{Verma07}
A.~K. Verma, P.~Brisk, and P.~Ienne, ``Rethinking custom {ISE} identification:
  a new processor-agnostic method,'' in \emph{Proceedings of the 2007
  international conference on Compilers, architecture, and synthesis for
  embedded systems}, ser. CASES '07, Salzburg, Austria, 2007, pp. 125--134.

\bibitem{Foggia01}
P.~Foggia, \emph{The {VFLib} Graph Matching Library}, 2nd~ed., Mar. 2001,
  http://amalfi.dis.unina.it/graph/db/vflib-2.0.

\bibitem{SPARK}
``The {SPARK} {HLS} tool homepage,'' 2003, http://mesl.ucsd.edu/spark/.

\bibitem{VCG}
G.~Sander, ``{VCG},'' 1996, http://rw4.cs.uni-sb.de/\~sander/html/gsvcg1.html.

\bibitem{Graphviz}
``Graphviz,'' 2010, http://www.graphviz.org.

\bibitem{CDFGtool}
``{CDFG} toolset,'' 2005, http://poppy.snu.ac.kr/CDFG/cdfg.html.

\bibitem{AGG}
``{AGG} transformation system,'' 2008, http://tfs.cs.tu-berlin.de/agg/.

\bibitem{Fisher04}
J.~A. Fisher, P.~Faraboschi, and C.~Young, \emph{Embedded Computing : A {VLIW}
  Approach to Architecture, Compilers and Tools}.\hskip 1em plus 0.5em minus
  0.4em\relax Morgan Kaufmann, December 2004, http://www.vliw.org/book/.

\end{thebibliography}

\end{document}